\documentclass{aa}
\usepackage{epsfig}
\usepackage{psfig}
\usepackage{graphicx}
\usepackage{txfonts}
\usepackage{lscape}
\newcommand{\chisq}{$\chi^2$}

\newcommand{\RX}{RX~J0822$-$4300}
\newcommand{\Pu}{Puppis$-$A}

\begin{document}

\title{X-Ray Observations of RX J0822-4300 and Puppis-A}
\author{C. Y. Hui \and W. Becker} 
\date{Received 10 June 2005 / Accepted 24 March 2006}
\institute{Max-Planck Institut f\"ur Extraterrestrische Physik, 85741 Garching bei M\"unchen, Germany}

\abstract{
 Based on observations with the X-ray observatories Chandra and XMM-Newton we present 
 results from a detailed spectro-imaging and timing analysis of the central compact 
 X-ray source RX J0822-4300 in the supernova remnant \Pu. The superior angular resolution 
 of Chandra allows for the first time to pinpoint the point source nature of this object down 
 to $0.59\pm0.01$ arcsec (FWHM) and to determine its position:  
 {RA=$08^{\rm h}21^{\rm m}57.40^{\rm s}$, Dec=$-43^{\circ}00^{'}16.69^{''}$ (J2000)}
 with sub-arcsecond accuracy. Spectral fits based on Chandra and XMM-Newton data provide 
 a tight constraint on the emission properties of \RX. Most of its X-ray emission seems 
 to be of thermal origin. A model spectrum consisting of two blackbody components with 
 $T_{1}\simeq 2.6\times10^{6}$ K, $T_{2}\simeq 5.0\times10^{6}$ K and $R_{1} \simeq 3.3$ km, 
 $R_2 \simeq 0.75$ km for the blackbody temperatures and the size of the projected 
 emitting regions, respectively, provides the best model description of its spectrum. 
 A search for X-ray pulsations from \RX\, revealed an interesting periodicity  
 candidate which, if confirmed, does not support a scenario of steady spin-down.

\keywords{pulsars: individual (RX J0822-4300)---stars: neutron--- supernovae: 
 individual (Puppis-A)---X-rays: stars} }

\maketitle
\section{Introduction}

 For many years, it has been generally believed that all young neutron stars have similar 
 properties as those observed in young rotation-powered pulsars, i.e.~emitting strongly 
 pulsed plus powerful plerionic radiation caused by non-thermal emission processes in the 
 neutron star's magnetosphere. Many recent observations of compact X-ray sources in 
 supernova remnants (SNRs), however, suggest that this picture is incomplete and no longer 
 justified. Apart from appearing as rotation-powered pulsars, it has been shown that there 
 are other manifestations of young neutron stars, e.g.~with no radio counterpart identified. 
 There are slowly rotating (P$\sim6-12\,\mbox{s}$) compact objects which possibly have an 
 ultra strong (B $\sim10^{14}-10^{15}\,\mbox{G}$) magnetic field. These neutron stars are 
 dubbed as magnetars, which include the {\em anomalous} X-ray pulsars (AXPs) 
 and the soft gamma-ray repeaters (SGRs),  
 depending on whether bursts of strong $\gamma$-ray emission is detected from them 
 (e.g., Mereghetti 1998; Thompson 2000). The other class of objects are the ``radio-quiet 
 neutron stars" (e.g.~Brazier \& Johnston 1999). Most of them were identified by their 
 high X-ray to optical flux ratios, others simply by their locations near to the expansion 
 centers of SNRs (e.g.~Becker \& Pavlov 2001, Kaspi et al.~2004), strongly suggesting that 
 they are indeed the compact stellar remnants formed in the supernova events. The group of 
 SNRs which are known to host a radio-quiet but X-ray bright central compact object is a 
 slowly growing one. Thanks to more sensitive X-ray observatories it currently includes 
 Cas$-$A (Tananbaum 1999), the Vela-Jr.~remnant (RX J0852.0$-$4622; Aschenbach 1998), RX 
 J1713.7$-$3946 (Pfeffermann \& Aschenbach 1996), RCW\,103 (Tuhoy \& Garmire 1980), 
 \Pu\,(Petre et al.~1982), PKS 1209-51/52 (Helfand \& Becker 1984) and Kes 79 (Seward et 
 al.~2003; Gotthelf et al.~2005). 
 
 The discovery of the X-ray point source, RX~J0822$-$4300, in \Pu\, was initially made 
 in one of the EINSTEIN HRI images of the SNR G260.4$-$3.4 (Petre et al.~1982). \RX\, 
 appeared in this data as a faint {\em X-ray feature}. With ROSAT, it became strongly 
 evident that RX~J0822$-$4300 is the compact stellar remnant which was formed in the SN
 event (Petre, Becker \& Winkler 1996; hereafter PBW96), although the positional offset 
 from the SNR's optical expansion center is 6.1 arcmin (cf.~Winkler \& Kirshner 1985, 
 Winkler et al.~1988). The age of \Pu, estimated from the kinematics of oxygen-rich 
 filaments is $\sim3700$ years. The remnant's kinematic distance estimated from an HI 
 study of the interstellar medium along the line of sight towards \Pu\, is $\sim 2.2 \pm 0.3$ 
 kpc (Reynoso et al.~1995; 2003). The space velocity of \RX\, required to travel  to its 
 observed position thus is $\sim 1000$ km/s. This is very high if compared with the mean 
 proper motion velocity observed in ordinary field pulsars but still comparable with what 
 is observed in several of the other young supernova/pulsar associations (Manchester et 
 al.~2005). 
   
 RX~J0822$-$4300 has not been detected as a radio pulsar.  Limiting radio flux densities 
 at 436 MHz, 660 MHz and 1520 MHz are 1.5 mJy, 1.3 mJy and 0.3 mJy, respectively (Kaspi et 
 al.~1996). For comparison, the limiting flux density in the Parkes Multi Beam  Survey 
 along the galactic plane was $\approx 0.2$ mJy (Manchester et al.~2001) and the typical 
 limiting sensitivity in deep searches for young radio pulsars in SNRs is 50 $\mu$Jy (Camilo 
 2003). Gaensler, Bock \& Stappers (2000) searched for a radio nebula around \RX\, with a 
 resolution of $27 \times17$ arcsec. Their non-detection of any extended plerionic radio 
 emission up to a scale of 30 arcmin prompted them to conclude that if \RX\, is a rotation-powered 
 pulsar, e.g.~with the radio beam not intersecting with the observer's line of sight, then 
 it must be less powerful than other typical young radio pulsars located in SNRs. All young  
 radio pulsars which are associated with a SNR have a spin-down power in excess  of 
 $\sim 10^{36}\,\mbox{ergs/s}$ and are observed to power a X-ray/radio bright pulsar-wind 
 nebula.  
 
 \RX\, has no optical counterpart down to a limiting magnitude of B $\gtrsim 25.0$ and 
 R $\gtrsim 23.6$ (PBW96). This limit yields an X-ray-to-optical flux ratio $f_{X}/f_{B} 
 \gtrsim 5000$ (PBW96). Together with the radio upper limits this rules out many types 
 of X-ray sources as a likely counterpart of \RX, except a neutron star.
 
 PBW96 fitted the ROSAT PSPC spectrum with a blackbody model and obtained  a temperature 
 of $(3.2\pm1.2)\times10^{6}$ K and a column density of $(4.1\pm0.2)\times10^{21}\,
 \mbox{cm}^{-2}$. The radius of the corresponding blackbody emitting area in their 
 fits is only $\sim 2$ km. Zavlin, Tr\"{u}mper, \& Pavlov (1999;  hereafter ZTP99) 
 tested whether a hydrogen atmosphere model could bring this result in better 
 agreement with the predictions of standard cooling models. They fitted a temperature 
 which is about half that found by PBW96 though with an increased radius of 10 km for 
 the emitting area. However, as atmosphere models are seen not to be in agreement with 
 the spectral fits from the cooling neutron stars Geminga, PSR B0656+14 and PSR B1055-52 
 (e.g.~De Luca et al. 2005),  the applicability of those models which in most cases 
 use non-magnetic opacities only, is restricted.
 
 EINSTEIN and ROSAT data do not show any evidence for short or long term flux variations. 
 Although a marginal detection of X-ray pulses at a period of $\sim75.3$ ms was claimed 
 by Pavlov, Zavlin, \& Tr\"{u}mper (1999; hereafter PZT99), it could not be confirmed so 
 far (Pavlov et al. 2002; Becker \& Aschenbach 2002).
 
 In order to put tighter constraints on the emission properties of \RX, various observations 
 with the new generation X-ray satellites XMM-Newton and Chandra were targeted to it in the 
 past few years. Making use of XMM-Newton's huge collecting power and high spectral resolution 
 as well as of Chandra's sub-arcsecond angular resolution we have performed a sensitive 
 broadband spectro-imaging analysis of \RX\, and its environment using all XMM-Newton and 
 Chandra data taken from this source so far. This is the subject of this paper which is 
 organized as follows. In \S2 we give a brief description of the relevant XMM-Newton and 
 Chandra observations. In \S3, we present the methods and results of our data analysis which 
 are discussed in \S4 in the context of a number of physical models for the nature of \RX. 

 \section{Observations}

 In total, five observations have been targeted with XMM-Newton and Chandra on \RX. All data 
 have been taken between December 1999 and  November 2001. We summarize the basic information 
 of these observations in Table \ref{obs_info} and give a more detailed description in the 
 following subsections.

 \subsection{XMM-Newton Observations}

 Two of the five data sets reported here were obtained with the {\bf E}uropean {\bf P}hoton 
 {\bf I}maging {\bf C}amera (EPIC) aboard XMM-Newton (Jansen et al.~2001). EPIC consists of 
 two Metal Oxide Semiconductor (MOS1/2) CCD detectors (Turner et al.~2001) of which half of
 the beam from two of the three X-ray telescopes is reflected to. The other two halves of the 
 incoming photon beams are reflected to a grating spectrometer (RGS) (den Herder et al.~2001). 
 The third of the three X-ray telescopes is dedicated to expose the EPIC-PN CCD detector solely 
 (Str\"uder et al.~2001). The April 2001 XMM-Newton observation (hereafter XMM1) was taken with 
 a total exposure time of $\sim 28.8$ ksec. The November 2001 observation (hereafter XMM2) had 
 an exposure time of $\sim 24.3$ ksec. The EPIC-PN CCD was operated in both observations in 
 small-window mode with a thin filter to block optical stray light. This data provide imaging,
 spectral and temporal information. All recorded events are time tagged with a temporal  
 resolution of 5.7 ms. The MOS1/2 CCDs were setup to operate in full-window mode with a medium 
 filter in the April 2001 observation and a thick filter in the November 2001 observation. The 
 MOS1/2 cameras provide imaging, spectral and timing information, though the later with a temporal 
 resolution of 2.6 s only. 
 
 For both XMM-Newton observations, the satellite was pointed to RA=$08^{\rm h}21^{\rm m}56^{\rm s}$ 
 and Dec=$-43^\circ 00' 19''$ [J2000]) which places \RX\, at the optical axis in the EPIC-PN CCDs. 
 The raw data from the EPIC instruments were processed with version 6.0.0 of the XMM Science Analysis 
 Software. Examining the raw data from the EPIC-PN CCD for both XMM1 and XMM2, we  did not find any 
 timing anomaly observed in many of the XMM-Newton data sets (cf.~Becker \& Aschenbach 2002; Kirsch 
 et al.~2004). This provides us with opportunities  for an accurate timing analysis. We created 
 filtered event files for the energy range 0.3 keV to 10 keV for all EPIC instruments. A small 
 fraction of X-ray events might be split between CCD pixels. In order to correct for this effect
 only those events were accepted for which the corresponding X-ray generated pattern was between 
 $0-12$ in MOS cameras and between $0-4$ in the EPIC-PN camera \footnote{For a detailed description 
 of EPIC event grade selection, please see the XMM-Newton Users' Handbook.}.
 We further cleaned the data by accepting only the good times when sky background was low and 
 removed all events potentially contaminated by bad pixels. The {\em effective} exposure times after 
 data cleaning are summarized in column 7 of Table \ref{obs_info}.

 In order to correct for the non-uniformity across the detector and the mirror vignetting, exposure map is needed to 
 rescale all parts of the image to the same relative exposure. This is created by using XMMSAS task EEXPMAP. 

 \RX\, is located in a patchy SNR environment. This makes the extraction of its source and background 
 spectrum difficult. In order to maximize the signal-to-noise ratio for \RX, we extracted its source
 spectrum from circles with 18 arcsec radii in both, the MOS1/2 and EPIC-PN cameras. About 70\% of all
 point source events are located within the selection region. Annular regions with radii between $20-35$ 
 arcsec, centered at \RX, were used to extract the background spectra. The background  corrected count 
 rates are listed in column 8 of Table \ref{obs_info}. Response files were computed for all data sets
 by using the XMMSAS tasks RMFGEN and ARFGEN.

\subsection{Chandra Observations}

 Three of the five data sets on \RX\, were taken with the Chandra satellite (e.g.~Weisskopf 2004). 
 One observation 
 was performed by using the Advanced CCD Imaging Spectrometer (ACIS; Burke et al.~1997) 
 whereas the other two exposures were done by using the High Resolution Camera (HRC; Zombeck 
 et al.~1995; Murray et al.~1997).  For the data reduction we used CIAO 3.0.2. 

 The Chandra HRC data were taken on 1999 December 21 and 2001 January 25 for HRC-I and 
 HRC-S, respectively, with \RX\, placed $\sim0.3$ arcmin off-axis. In order to determine the
 event positions accurately, we started the analysis with level-1 event files and corrected 
 for the tap-ringing distortion in the HRC event position reconstruction. Apart from this, we 
 also performed the de-gap correction to the event files so as to compensate the systematic 
 errors introduced in the event positions by the algorithm used to determine  the centroid of 
 the charge cloud exiting the rear micro-channel plate of the HRC. Furthermore, aspect offset 
 was corrected for the event files. 
  
 The ACIS observation was performed on 2000 January 1 using the front-illuminated (FI) ACIS-S2 
 chip with a frame time of 0.84 s. \RX\, is located $\sim 2.7$ arcmin off-axis in this ACIS 
 observation. In order to correct for possible pileup effects, we started our analysis again 
 with level-1 files as those have preserved a number of source events which could have been 
 misidentified as afterglows of cosmic ray events in the standard processing of level-2 data 
 (cf.~Davis 2002). The sub-arcsecond resolution of Chandra allows to extract the counts for 
 the spectral analysis from a circle with radius 2.6 arcsec (encircled energy $\sim99\%$ for 
 on-axis point sources). This selection radius minimizes the contamination from the supernova 
 background emission. An annular region with radii between $\sim 2.6-5.3$ arcsec, centered at 
 \RX\, was chosen to extract the background spectrum.  Response files were created using the
 tools MKRMF and MKARF of CIAO. The background  corrected HRC and ACIS count rates of \RX\, 
 are given in column 8 of Table \ref{obs_info}. 

 From the ACIS-S2 and XMM-Newton data we found that the energy of the central source peaks 
 at $\sim1.5$ keV. With the peak energy and for the off-axis angle of $\sim 0.3$ arcmin we 
 extracted the desired point spread function (PSF) model images from CALDB 2.26 standard 
 library files (F1) by interpolating within the energy and off-axis angle grids by using 
 CIAO tool MKPSF. Exposure maps for the corresponding images were generated by the tool 
 MKEXPMAP.

\section{Data Analysis}

\subsection{Spatial Analysis}

\begin{figure*}
\centerline{\psfig{figure=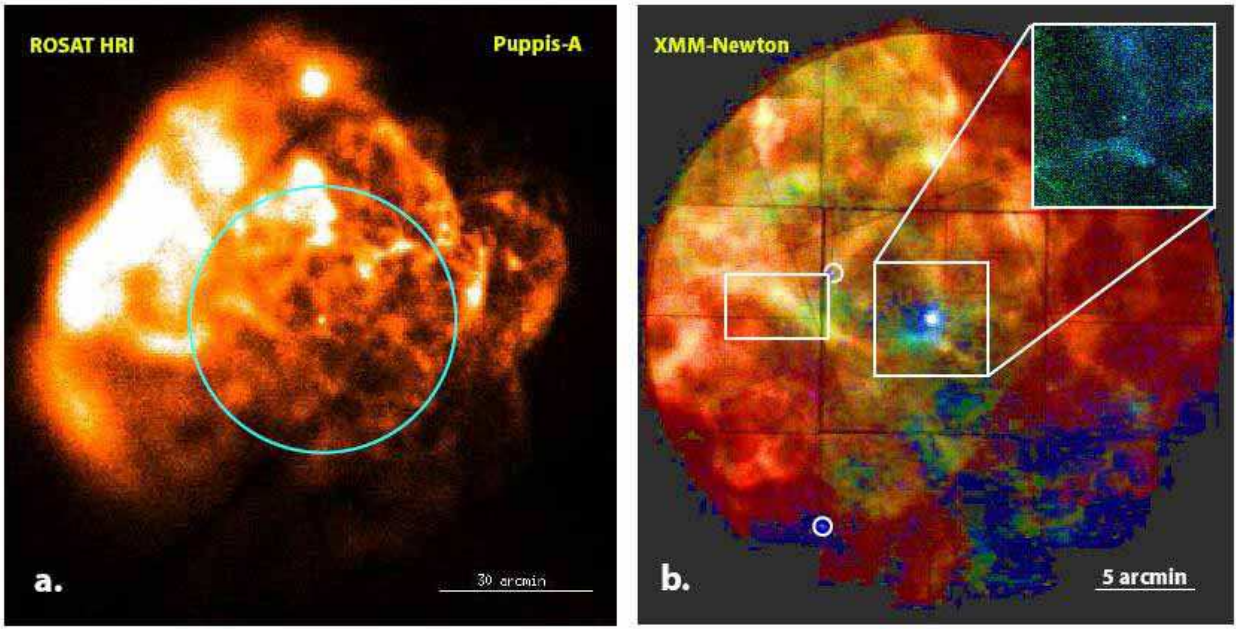,width=18cm,clip=}}
\caption{{\bf a.}~Composite ROSAT HRI image of the \Pu\, supernova remnant. The blue ring
indicates the 30 arcmin central region which has been observed by XMM-Newton in
April and November 2001.
{\bf b.}~XMM-Newton MOS1/2 false color image of the inner 30 arcmin 
central region of \Pu\, (red: $0.3 - 0.75$ keV, green: $0.75 - 2$ keV and blue: $2-10$ keV). 
The central source is \RX. The inset shows the squared region as observed by the 
Chandra HRC-I. It is interesting to note that the region around \RX\ comprise mainly 
hard X-ray photons. The left white box indicates the region of the 
swirl-like structure interpreted by Winkler et al.~(1989) as a second supernova in \Pu. 
The location of the northern and southern hard X-ray point sources are indicated by circles. 
The binning factors in the XMM and Chandra images are 4 arcsec and 0.6 arcsec, 
respectively. Adaptive smoothing with a Gaussian kernel of $\sigma<1$ pixel 
has been applied to the XMM-Newton and Chandra images. Top is north and left is east.} \label{rgb}
\end{figure*}

  Composite images of the supernova remnant \Pu\, and its central region around \RX,
  as seen by the ROSAT HRI, by XMM-Newton's MOS1/2 CCDs and by the Chandra HRC-I, are shown in  
  Figure \ref{rgb}. \RX\, is located at the center of these images. From Figure \ref{rgb}b 
  it can be seen that the hardest X-ray emission in the remnant is mainly contributed by the 
  central compact object \RX. Apart from this, we observed two more hard X-ray sources in the 
  XMM-Newton MOS1/2 images. Their locations as indicated by circles in Figure \ref{rgb}b are 
  RA$=08^{\rm h} 22^{\rm m} 26.70^{\rm s}$, Dec$= -43^\circ 10' 25.99''$ (J2000) for the source located 
  in the south and RA$=08^{\rm h} 22^{\rm m} 24.25^{\rm s}$, Dec$= -42^\circ 58' 00.82''$ (J2000)   
  for the northern source.  
  The location of the northern source is close to the region which was suggested  by Winkler et 
  al.~(1989) to be a second supernova within \Pu. These authors have observed an unusual swirl-like 
  structure in optical images and interpreted this as a possible second supernova remnant. The center 
  coordinate of this structure is at about RA$=08^{\rm h} 22^{\rm m} 39^{\rm s}$, 
  Dec$= -42^\circ 59' 41''$ (J2000). The left box in Figure \ref{rgb}b illustrates the field of 
  view in their observations. The angular separation between the northern hard X-ray source and the 
  center of the swirl-like structure is $\sim3.2$ arcmin. From the spectral analysis of optical 
  filaments, Winkler et al.~(1989) estimated that the kinematic age of the proposed second SNR is 
  $< 800$ years. If the northern hard X-ray source is correlated with this structure and this age 
  estimate is correct it would require a space velocity $> 2000$ km/s (for an assumed distance of
  2.2 kpc) in order to travel to its observed location. An association thus would be unlikely due 
  to this high space velocity. For the southern hard X-ray source its correlation with  \Pu\ is 
  unspecified though most likely this is a background source. The photon statistics does not 
  support a detailed spectral analysis for these two sources. 

\begin{figure*}
\centerline{\psfig{figure=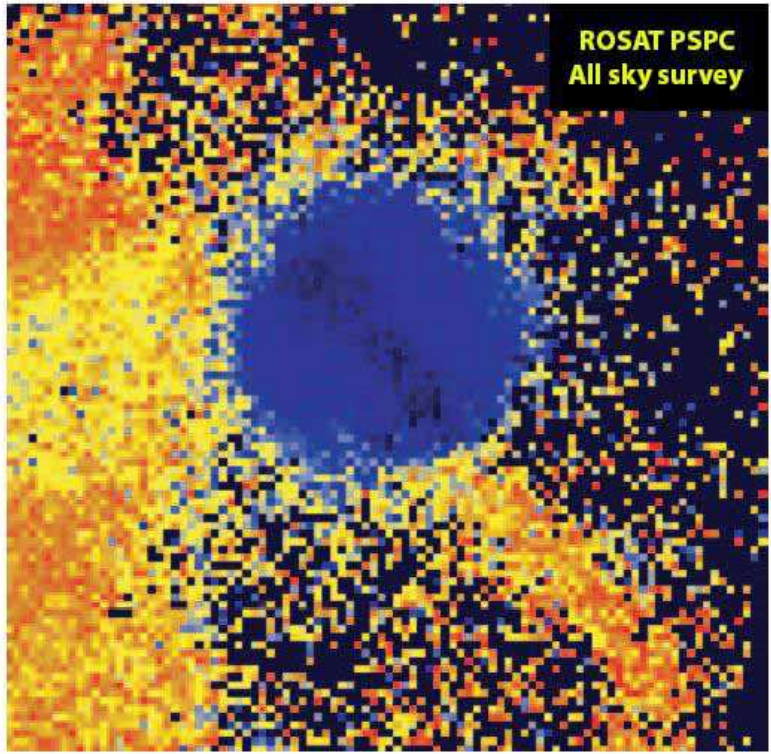,width=18cm,clip=}}
\caption[]{False color image of \Pu\, and parts of the Vela supernova remnant as seen
in the ROSAT all-sky survey. \Pu\, appears solely in blue color (harder X-rays) while 
the soft emission from the Vela supernova remnant is represented by the red and yellow colors. 
The image clearly demonstrates the existence of the absorption belt which crosses \Pu\, from
the south-west to north-east direction and which is associated with rim emission from the 
Vela supernova remnant.} 
\label{rosat_survey}
\end{figure*}

  The XMM-Newton MOS1/2 false color image (Figure \ref{rgb}b) demonstrates nicely that the 
  south-western part of the remnant as well as the region near \RX\ comprise mainly hard X-ray 
  photons. This is different from other parts of the image which consist of soft X-rays from 
  the hot supernova ejecta. As \Pu\ is located at the edge of the Vela supernova remnant 
  (distance $\sim0.25$ kpc), and is located behind it, we speculate that there is intervening 
  absorbing material from Vela along the line of sight which absorbs most of the soft X-ray photons
  of the south-western part of \Pu. This view is supported by Figure \ref{rosat_survey} which 
  shows a belt of absorbing material crossing the whole \Pu\, supernova remnant from the
  south-western to the north-eastern direction (Aschenbach 1994 and discussion therein).

\begin{figure*}
\centerline{\psfig{figure=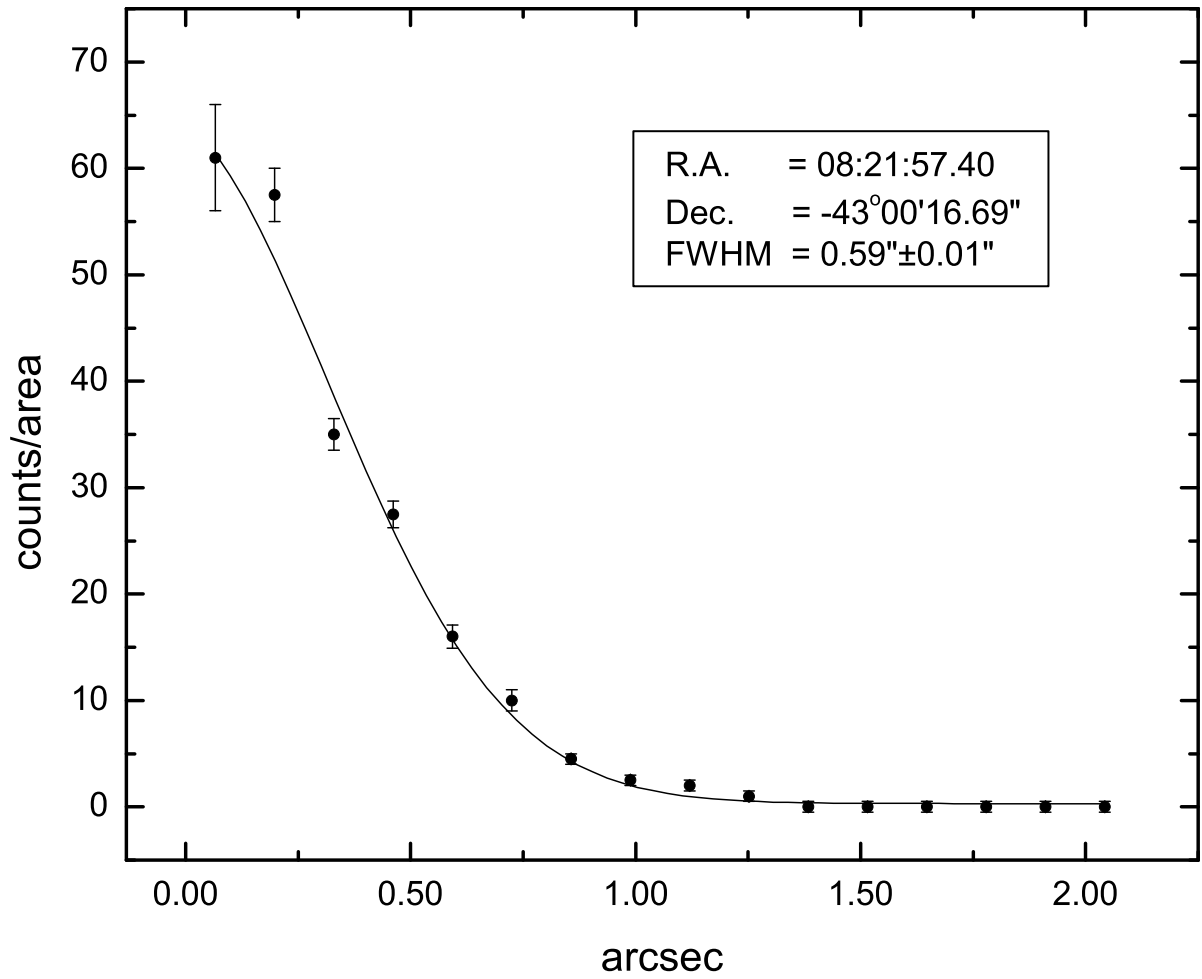,width=18cm,clip=}}
\caption{Chandra HRC-I brightness profile for RX J0822-4300. The solid curve represents the
best-fit Gaussian model with the modeled PSF at 1.5 keV as a convolution kernel.} \label{chandra_psf}
\end{figure*}

  The high resolution X-ray image from HRC-I allows for the first time to examine the spatial 
  nature of \RX\, with sub-arcsecond resolution. However, we have found that the full width half 
  maximum (FWHM) of the point spread function (PSF) ($\sim$ 0.4 arcsec) generated from the library 
  files is narrower than expected. This can be ascribed to the fact that the PSF library files are 
  derived by a ray-tracing program instead of obtained directly from the calibration data. Due to this caveat, 
  it is legitimate to fit the image with the convolution of a 2-dimensional Gaussian function instead 
  of a delta function. The radial profile of RX J0822-4300 is depicted in Figure \ref{chandra_psf}. 
  The solid curve represents the best-fit Gaussian model with the modeled PSF at 1.5 keV  as a 
  convolution kernel. The best-fit results yield a FWHM of $0.59 \pm0.01$ arcsec which is very close 
  to the expected width of the Chandra PSF. This result appears  as the first evidence for the point 
  source emission nature of \RX. Moreover, we were  also able to narrow down the position of this 
  compact object to the smallest region that has never been obtained before. The best-fit gives us 
  a position for \RX\, which is RA$= 08^{\rm h} 21^{\rm m} 57.40^{\rm s}$ and 
  Dec$= -43^\circ 00' 16.69''$ (J2000). The statistical error of the position introduced by the image-fitting
  is found to be 0.01 arcsec ($1-\sigma$). The predominant uncertainty of the source position is 
  given by the finite width of the PSF ($\sim$ 0.5 arcsec) and the average pointing accuracy of the 
  satellite ($\sim$ 0.6 arcsec). 
  The same position is obtained from the analysis of the HRC-S data. 
  The position and point source character of \RX, 
  deduced by using Chandra, are in agreement with what we found in the XMM-Newton data 
  (PSF $\sim 5$ arcsec FWHM).

\begin{figure*}
\centerline{\psfig{figure=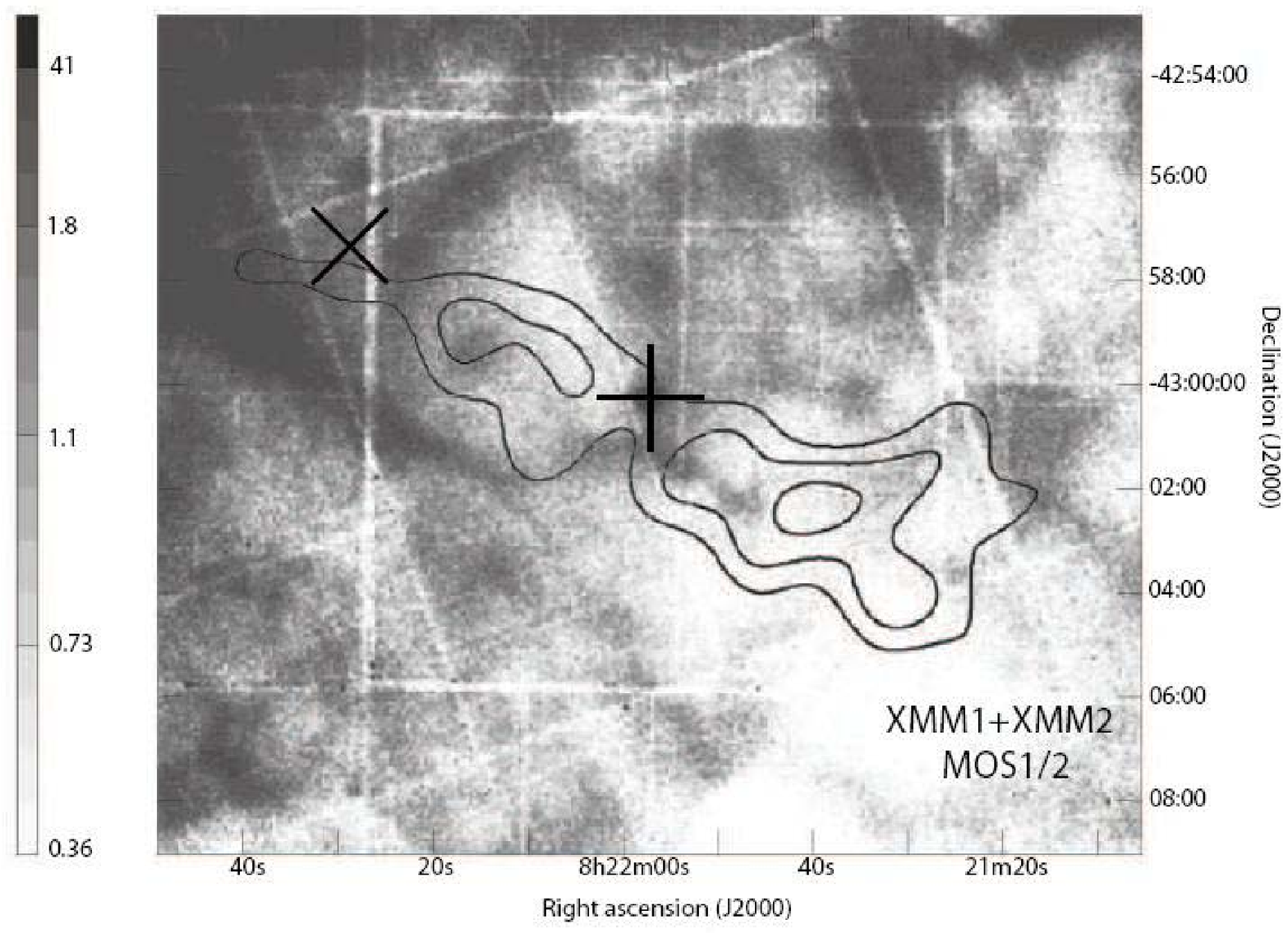,width=15cm,clip=}}
\centerline{\psfig{figure=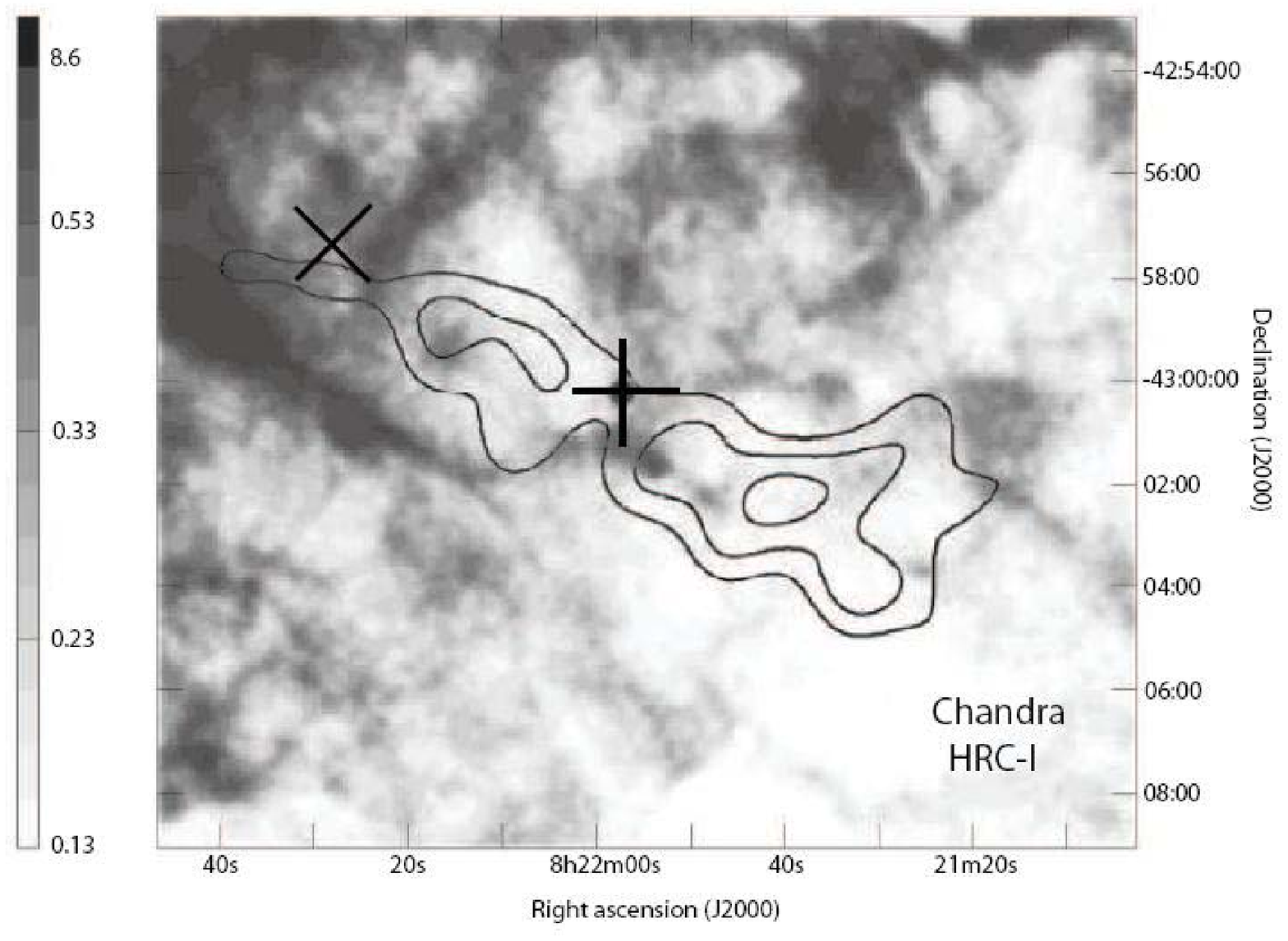,width=15cm,clip=}}
\caption{The contours depict a double-lobed feature of reduced $\lambda21$-cm line emission (from Reynoso et al. 2003) superimposed
on fluxed image of XMM-Newton MOS1/2 ({\it upper}) and Chandra HRC-I ({\it lower}) in gray scale. 
The cross shows the optical expansion center of \Pu\, as calculated by Winkler
et al. (1988).} \label{radio_HI}
\end{figure*}

  From observations of the neutral hydrogen surrounding \RX, Reynoso et al.~(2003) found 
  a depression in the $\lambda21-$cm line emission near to \RX. According to their 
  interpretation this structure could be connected to the compact stellar remnant because
  of its symmetric appearance as well as because of its alignment with the 
  remnants optical expansion center and the position of \RX. In order to 
  search whether there is an X-ray structure near to \RX\, which correlates 
  with this radio structure we have overlaid the radio contours from 
  Reynoso et al.~(2003) on the XMM-Newton and Chandra HRC-I image (cf.~Figure 
  \ref{radio_HI}). No clear correlation between the radio and  X-ray structures 
  is seen, though the patchy supernova environment makes any conclusion uncertain.

\subsection{Spectral Analysis}

  We estimated the effects of pileup in both XMM1 and XMM2 data by using the XMMSAS task EPATPLOT. 
  Our results showed that all the EPIC data were not affected by CCD pileup. Using the spectral  
  parameters of \RX\, inferred from XMM-Newton, we estimated with the aid of PIMMS (version 3.6a) 
  that the ACIS-S2 data are piled-up by a fraction of $\sim11\%$. We applied adequate correction
  by incorporating a pileup model in the spectral fitting (Davis~2001). Chandra data were also 
  corrected for the degradation of quantum efficiency. 

  In order to constrain the spectral parameters tightly, we fitted XMM1, XMM2, as well as the
  ACIS-S2 data simultaneously. In order to obtain spectra from different observations and 
  instruments with compatible significance, 
  the energy channels were grouped dynamically with respect to the 
  photon statistics in the analyzed data sets. For the MOS1/2 data of XMM2, we grouped the data 
  to have at least 50 counts per bin. For the MOS1/2 data of XMM1 as well as the ACIS-S2 data 
  we applied a grouping so as to have at least 100 counts per spectral bin. For the EPIC-PN data 
  from XMM1 and XMM2 a grouping of 200 cts/bin was used. All spectral fits were performed in the 
  $0.3-10$ keV energy range by using XSPEC 11.3.1.

  Various model spectra like single blackbody, double blackbody, power-law, combinations of 
  blackbody and power-law, broken power-law, as well as thermal bremsstrahlung were fitted to 
  the data. Independent of the fitted spectral models we found that the fits improve if the 
  spectrum extracted from the Chandra data was not included in the analysis. Since the spectra 
  extracted from XMM-Newton data supersedes the Chandra data in photon statistics, we excluded 
  the later from the spectral analysis without loss of generality. The parameters of all fitted 
  model spectra are summarized in Table 2. The quoted errors are conservative and are 
  1$\sigma$ for 2 parameters of interest for single component spectral models and for 
  3 parameters of interest for multi-component model.

  Fitting the spectral parameters of \RX\, as inferred from ROSAT PSPC data by PBW96 we found 
  that these parameters ($T=3.2\times10^{6}$K, $N_{H}=4.1\times10^{21} \mbox{cm}^{-2}$, $R=2$ km) 
  yield no acceptable description of the XMM-Newton observed spectrum ($\chi^{2}_{\nu}=21.17$ for 
  467 dof). In general, spectral fitting with a single component blackbody or power-law  model did
  not model the data beyond $\sim3$ keV (cf.~Table 2). Testing multi-component models 
  we found that a two component blackbody with $N_{H}=4.54^{+0.49}_{-0.43} \times10^{21}
  \mbox{cm}^{-2}$, $T_{1}=2.61^{+0.30}_{-0.26}\times10^{6}$K, $T_{2}=5.04^{+0.28}_{-0.20}
  \times10^{6}$K and $R_{1}=3.29^{+1.12}_{-0.74}$ km, $R_{2}=0.75^{+0.12}_{-0.15}$ km
  for the blackbody temperatures and emitting areas, respectively, yields the best description 
  of the observed spectrum.  The reduced-$\chi^{2}$ of this fit is 1.20 for 465 dof. We note 
  that the apparent deviation of the reduced-$\chi^{2}$ from one, indicating an acceptable fit, is 
  due to the fact that the data from different instruments and different epochs are modeled 
  simultaneously. The benefit of combining all spectral data in simultaneous fits is the higher 
  photon statistics and thus the ability to better discriminate between competing model spectra. 

\begin{figure*}
\centerline{\psfig{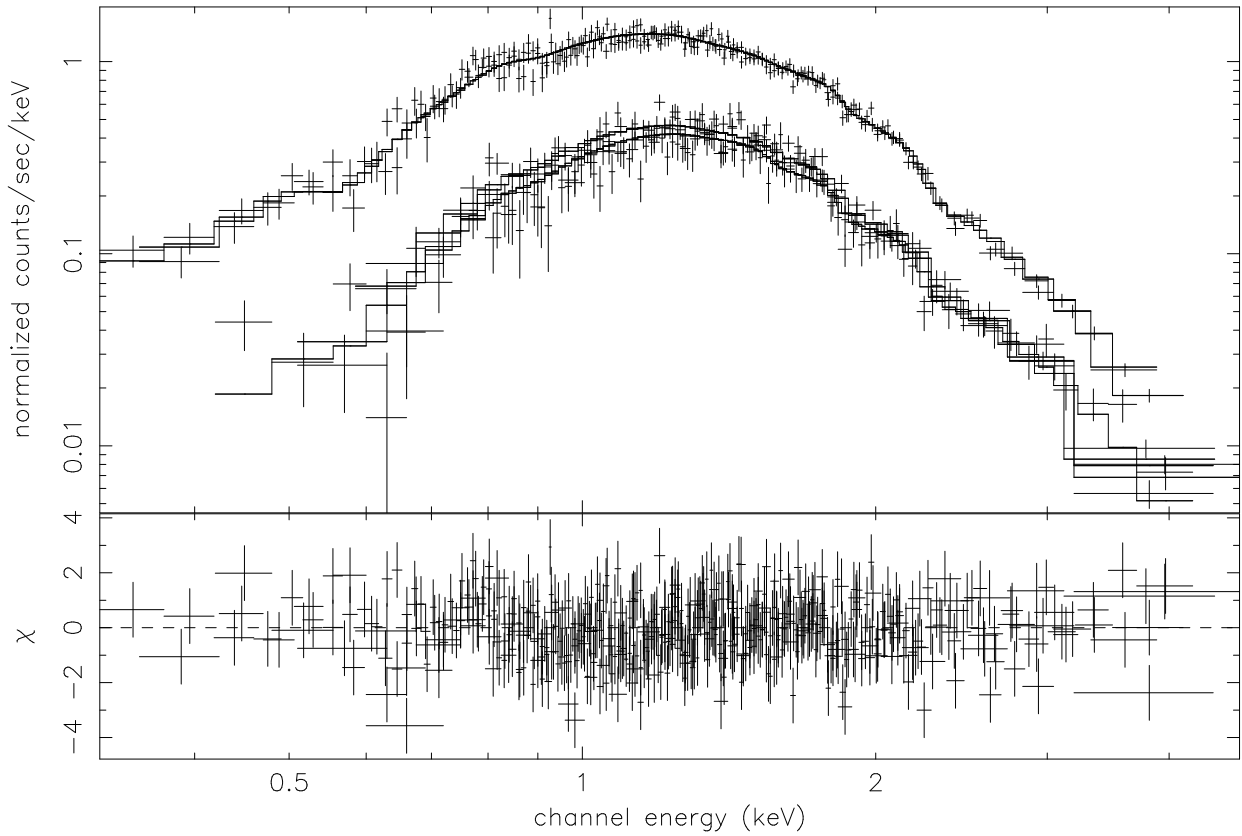}}
\caption{Energy spectrum of \RX\, as observed with the EPIC-PN (upper spectra)
and MOS1/2 detectors (lower spectra) and simultaneously fitted to an absorbed 
two component blackbody model ({\it upper panel}) and contribution to the \chisq\, fit statistic
({\it lower panel}).} \label{RX_spectrum}
\end{figure*}

\begin{figure*}
\centerline{\psfig{figure=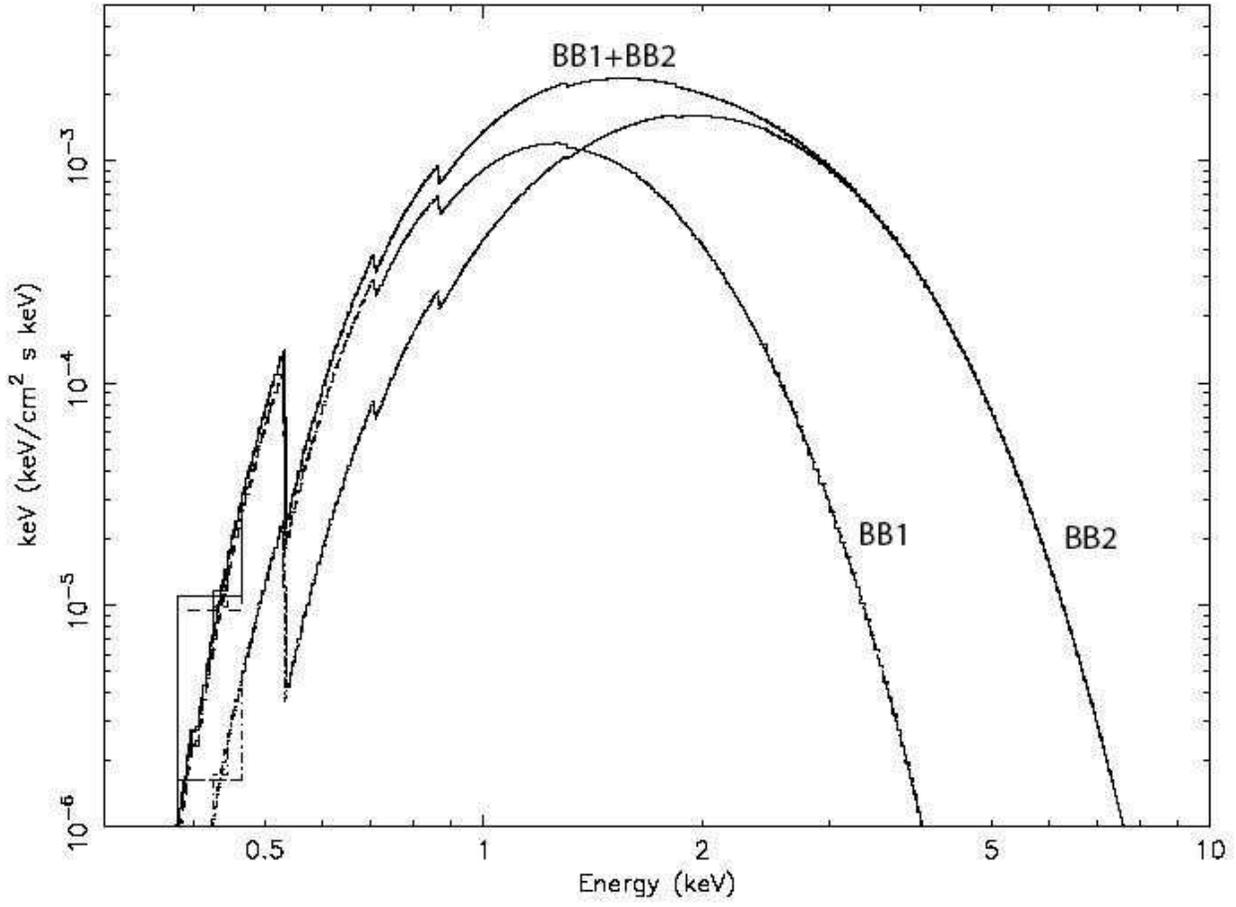,width=18cm,angle=0,clip=}}
\caption{Components and combined model of double blackbody fitted to the spectral data of \RX.}\label{model}
\end{figure*}

\begin{figure*}
\centerline{\psfig{figure=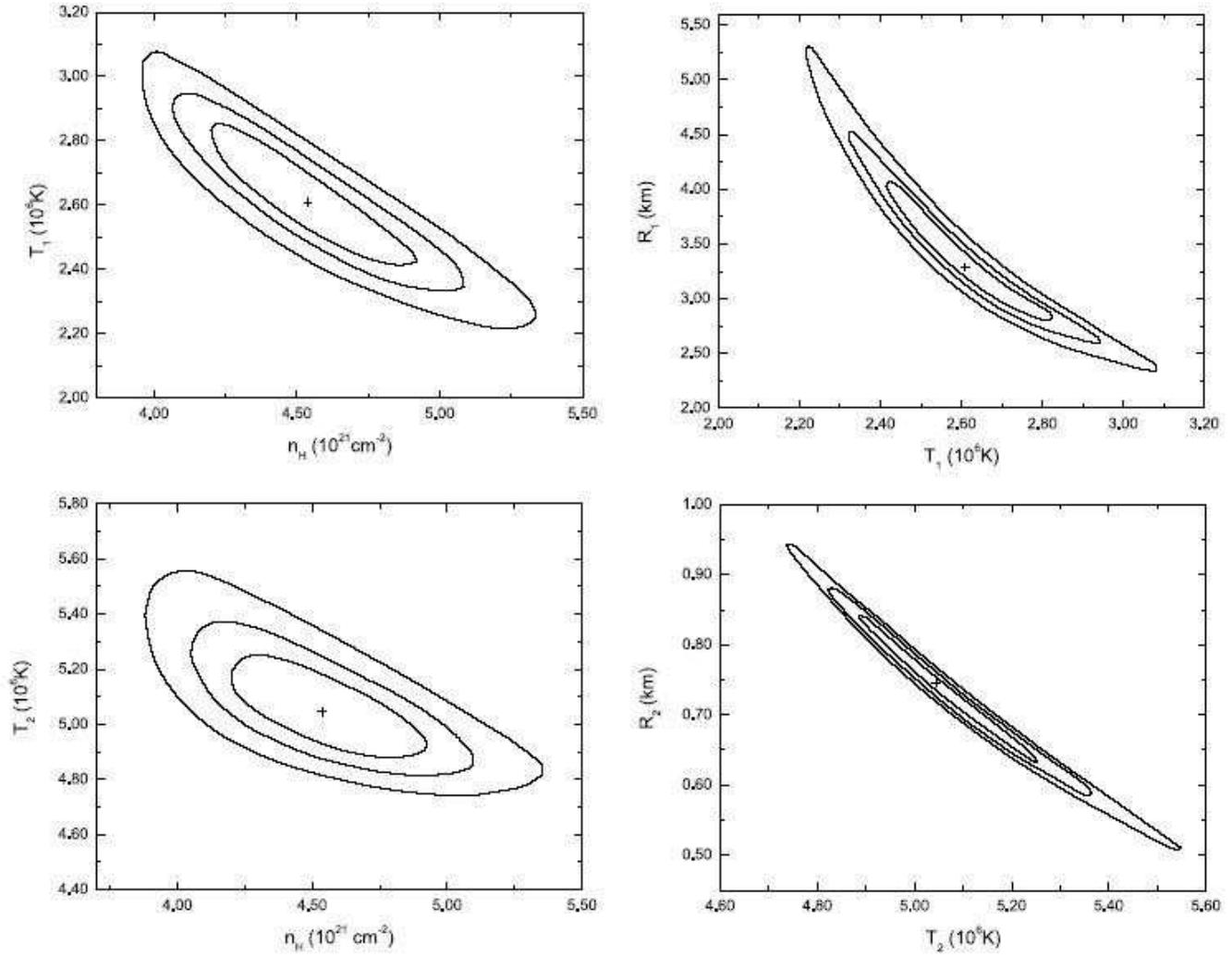,width=18cm,clip=}}
\caption{1$\sigma$, 2$\sigma$ and 3$\sigma$ confidence contours for the double blackbody 
fit to the X-ray spectrum of \RX.}\label{contours}
\end{figure*}

  Figure \ref{RX_spectrum} and \ref{model} shows the spectral fit for an absorbed double blackbody model and 
  the corresponding spectral components respectively. 
  In order to properly constraint the parameter space for the best-fitting model, we calculated 
  the contour plots in the $T_{1}-R_{1}$, $T_{1}-N_{H}$ and $T_{2}-R_{2}$, $T_{2}-N_{H}$ planes, 
  respectively. These plots are depicted in  Figure \ref{contours}. For a consistence check we 
  modeled the Chandra ACIS-S2 spectrum with the best fitting double blackbody model and found 
  all parameters in agreement with those fitted for the XMM-data. 

  Both $R_{1}$ and $R_{2}$ inferred from the double blackbody fit are inconsistent with the 
  size of a canonical neutron star (i.e.~$R\sim 10$ km). It is therefore instructive to redo 
  the fitting for this model with $R_{1}$ fixed at 10 km. This model still yields acceptable 
  values of $N_{H}=6.38^{+0.21}_{-0.13}\times10^{21}\mbox{cm}^{-2}$, 
  $T_{1}=1.87^{+0.02}_{-0.02}\times10^{6}$K, $T_{2}=4.58^{+0.03}_{-0.07}\times10^{6}$K, and   
  $R_{2}=1.09^{+0.04}_{-0.04}$ km with $\chi^{2}_{\nu}=1.28$ (for 466 dof) only a slightly 
  larger than leaving $R_{1}$ unconstrained.

  For a model combining a blackbody and a power-law, the goodness-of-fit is compatible with 
  that for the double blackbody model ($\chi^{2}_{\nu}=1.21$ for 465 dof). The inferred 
  slope of the power-law component is $\Gamma=4.67^{+0.14}_{-0.05}$. Although this 
  is steeper than the photon-index, $\Gamma=1-3$, observed for rotation-powered pulsars 
  (cf.~Becker \& Tr\"{u}mper 1997), the model cannot be rejected simply based on this 
  as it is not a priori applicable for central compact objects. 
  However, the column density is much higher than the expected level. 
  When $N_{H}$ is fixed to $4\times10^{21}\mbox{cm}^{-2}$, which is consistent with the values 
  obtained by PBW96, ZTP99 and Winkler et al. (1981), it results in a parameter set of 
  $\Gamma=2.51^{+0.11}_{-0.13}$, $T=3.80^{+0.06}_{-0.05}\times10^{6}$K, 
  $R=1.59^{+0.06}_{-0.06}$ km, though with a large $\chi^{2}_{\nu}$ of 1.42 for 466 dof.

  It is necessary to examine whether a broken power law model can describe the spectra. 
  This implies a purely non-thermal emission with spectral steepening at high energy after 
  an energy break which is due to the deficit of energetic emitting charged particles. 
  From Table 2 it is obvious that the broken power law model does not yield any photon index 
  that is consistent with that of a typical pulsar. We also fitted the data with a thermal 
  bremsstrahlung model which physically implies that the central compact object would be 
  surrounded by a hot plasma. From the normalization constant inferred from the spectral
  analysis, we can calculate the extent of the plasma. Following Iaria et al. (2001), we assume
  the bremsstrahlung normalization to be $N_{bremss}=3.02\times10^{-15}N_{e}^{2}V/4\pi D^{2}$,
  where $D$ is the distance to the source in cm, $N_{e}$ is the electron density ($\mbox{cm}^{-3}$),
  and $V$ is the volume of the bremsstrahlung emitting region. 
  Assuming $N_{e}$ is comparable with the average density $\sim 1 \mbox{cm}^{-3}$ of \Pu\ 
  (Petre et al. 1982), the radius of the assumed spherical
  emitting region is  estimated to be $\sim 2.7$ pc for an adopted distance of 2.2 kpc. 
  This implies that the source should be
  extended (at a level of $\sim 4.2$ arcmin assuming a distance of 2.2 kpc), in contradiction
  to the results from the spatial analysis.
  
\begin{figure*}
\centerline{\psfig{figure=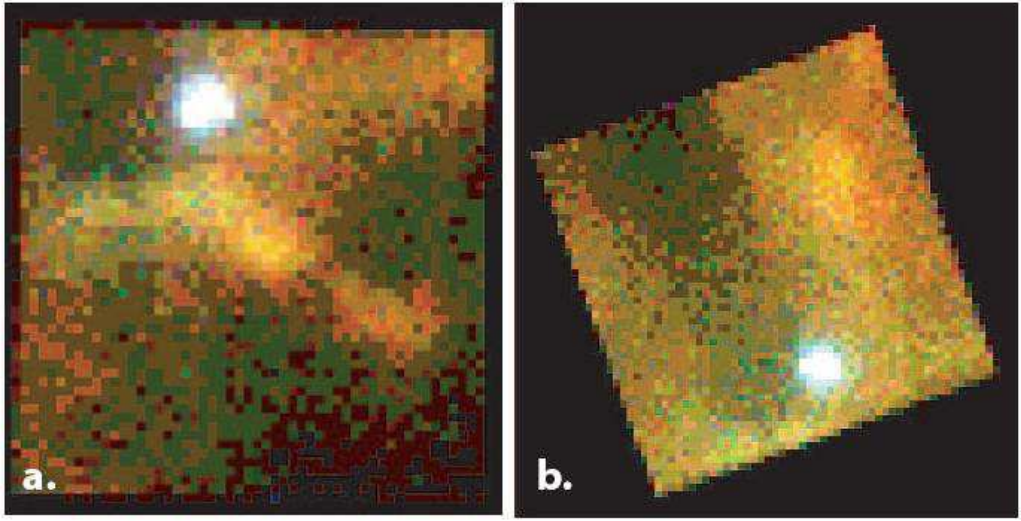,width=18cm,clip=}}
\caption[]{False color image (red: $0.3 - 0.75$ keV, green: $0.75 - 2$ keV and blue: $2-10$ keV)
of the $4.4' \times 4.4'$ region around RX J0822-4300 as seen by XMM-Newton's EPIC-PN detector
during the observations in April 2001 ({\bf a.}) and November 2001 ({\bf b.}). The rim emission
from \Pu\, near to the location of \RX\, is visible very well in {\bf a.} Top is north and
left is east.} \label{epic_pn}
\end{figure*}

\begin{figure*}
\centerline{\psfig{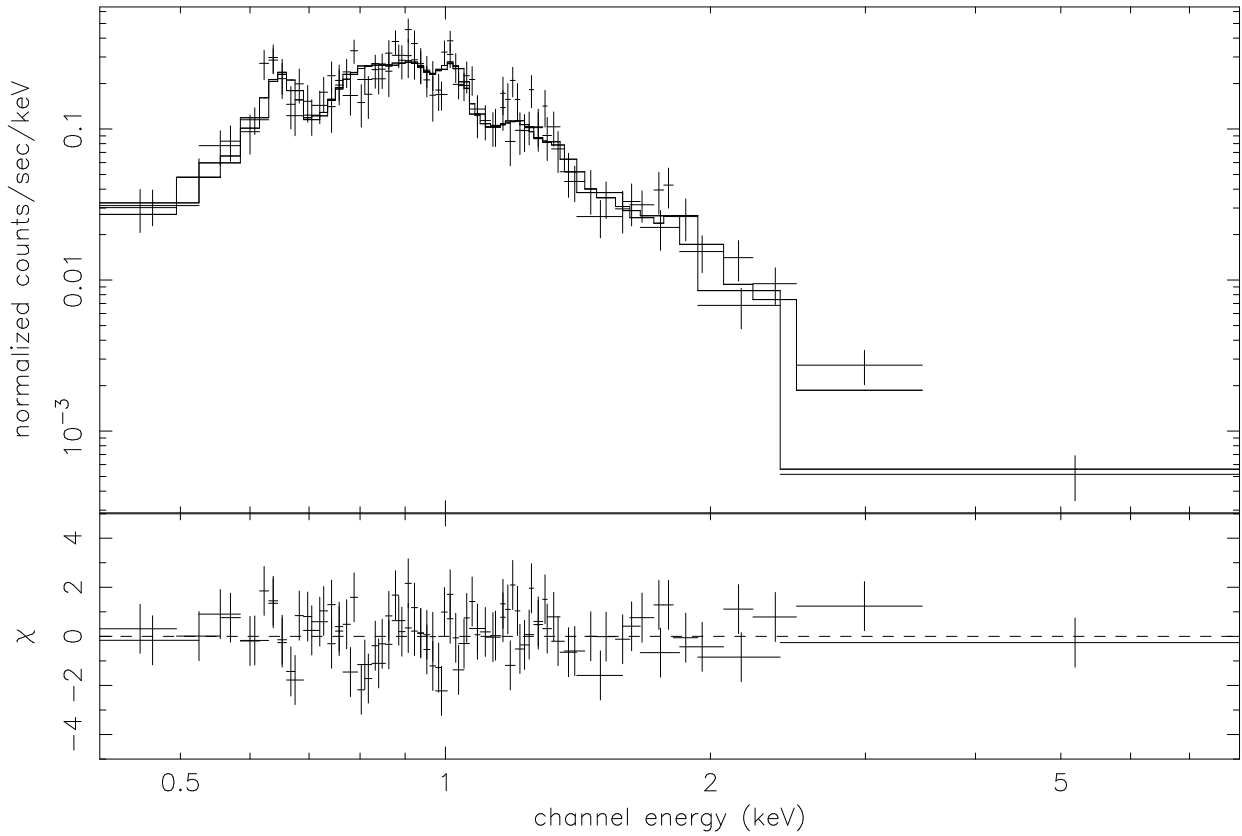}}
\caption{Energy spectrum of the rim emission underneath \RX\, as observed in April 2001 with the EPIC-MOS1/2 detector
and simultaneously fitted to an absorbed non-equilibrium ionization collisional plasma model ({\it upper panel}) 
and contribution to the \chisq\, fit statistic ({\it lower panel}).} \label{rim_spectrum}
\end{figure*}

  From both, the XMM-Newton MOS1/2 and the Chandra HRC-I images, some faint and diffuse hard 
  X-ray emission around \RX\, seems to be present (cf.~Figures \ref{rgb} \& \ref{epic_pn}). 
  Its nature can be determined by examining its spectrum. We extracted the events in the X-ray 
  filament near to \RX\ from the MOS1/2 cameras of XMM1 from a 80 arcsec$\times$30 arcsec box centered 
  at RA=$08^{\rm h}21^{\rm m}57.077^{s}$, Dec=$-43^{\circ}01'15".42$ (J2000). We found that its 
  spectrum is consistent with an absorbed non-equlibrium ionization collisional plasma model 
  (XSPEC model: VNEI) with goodness-of-fit of $\chi^{2}=151.08$ for 142 dof. 
  The energy spectrum as fitted to this model spectrum 
  is displayed in Figure \ref{rim_spectrum}. Line emission is easily recognized in this plot. 
  The most obvious feature is the O VII and O VIII line complex at 0.662 keV and 0.651 keV 
  respectively. 
  Parameters inferred from the best-fitted model are the column density 
  $N_{H}=3.70^{+0.12}_{-0.12}\times10^{21}\mbox{cm}^{-2}$, the plasma temperature 
  $T=7.62^{+0.10}_{-0.21}\times10^{6}$K, the ionization timescale $\tau=2.33^{+0.15}_{-0.11}
  \times10^{11}\mbox{s cm}^{-3}$, and the metal abundances with respect to the solar values 
  (O: $17.83^{+1.58}_{-1.56}$, Ne: $4.00^{+0.67}_{-0.68}$, Si: $2.25^{+1.33}_{-1.32}$, 
  S: $6.58^{+4.85}_{-4.81}$, Fe: $2.24^{+0.20}_{-0.21}$) (quoted errors are $1\sigma$ for 2 
  parameters of interest). These parameters imply a relative abundance ratio 
  O:Fe to be about $6-9$ times its solar value which strongly suggest an enhancement of oxygen in \Pu. 
  This is in agreement with the conclusion drawn by Canizares \& Winkler (1981). 
  For the other elements, including H, He, C, Mg, Ar, Ca and Ni, we do not find any sign of 
  enhancement and their abundances are in agreement with the solar values. 
  We have performed the spectral fitting with different selected backgrounds. Provide that 
  the backgrounds are selected from low count regions, all the best-fitted values are found to
  be within the quote $1\sigma$ errors above. Since the remnant 
  environment is patchy and inhomogeneous, abundance ratios from different regions are not expected 
  to be comparable. A detailed modeling of the variation of chemical abundance is beyond the scope 
  in this paper. A further detailed analysis of \Pu\ is in preparation and will be published elsewhere. 
 
  Although the rim emission appears to be a part of the structure of \Pu, we also exam whether there is 
  any non-thermal contribution in the emission by adding a Crab-like power-law component 
  (i.e. with photon index of 2) in the spectral fit. The additional component does not improve the goodness-of-fit at all 
  ($\chi^{2}=150.95$ for 141 dof). A $3\sigma$ upper bound of the power-law model normalization is estimated to be 
  $6\times10^{-5}$ photons keV$^{-1}$ cm$^{-2}$ s$^{-1}$. This implies a limiting flux of non-thermal plerionic emission, 
  if any, to be $3.06\times10^{-13}$ ergs cm$^{-2}$ s$^{-1}$ and $2.88\times10^{-13}$ ergs cm$^{-2}$ s$^{-1}$ in 
  0.1$-$2.4 keV and 0.5$-$10 keV respectively. 


\subsection{Timing Analysis}

\subsubsection{Search for long-term variabilities}

  To check whether the energy fluxes measured from \RX\, by XMM-Newton and Chandra are 
  consistent with each other or whether there are significant long-term deviations observed 
  in the different data sets we computed the flux for the best fitting double blackbody model 
  from all available data. In order to compare the XMM-Newton and Chandra results with the   
  existing ROSAT flux we restricted this computation to the energy range $0.1-2.4$ keV. As 
  shown in Table \ref{flux}, all observed energy fluxes, from ROSAT to XMM-Newton are consistent 
  with a constant energy flux of $f_x(\mbox{0.1-2.4 keV}) \sim 3\times 10^{-12}$ ergs cm$^{-2}$ 
  s$^{-1}$. 
  The same conclusion can be drawn from comparing the broadband fluxes from Chandra, XMM1 
  and XMM2. The observed flux of these three observations are found to be $\sim 4\times 10^{-12}$ 
  ergs cm$^{-2}$ s$^{-1}$ in the range of 0.5$-$10 keV. 
  

\subsubsection{Search for coherent pulsations}

  Although the lack of long-term variability and the spectral analysis eliminates some highly 
  improbable models and hence helps us to put constraints on the properties of the central 
  compact object, the most strong argument that this object is indeed a neutron star would 
  come from the detection of X-ray pulsations. Since the small-window mode was setup for the
  EPIC-PN  camera in both XMM1 and XMM2 observations, the 5.7 ms temporal resolution of this 
  data is sufficient to search for coherent short-term pulsations.  

  The arrival times in both event files were barycentric corrected using the XMMSAS task BARYCEN.
  In order to minimize the systematic errors induced in the barycentric correction, we use 
  the position inferred from the Chandra HRC-I image fitting (c.f.~\S3.1) for correcting the 
  arrival times in both data sets. The initial period searches were performed by applying a 
  fast Fourier transformation (FFTs) on both sets of photon arrival times separately. The 
  advantage of having multiple data sets from different epochs supports to cross-check 
  any periodicity candidates easily and prevents wrong identifications. FFTs were calculated 
  for each time series with 20 different binnings. Searches in the frequency domain were limited 
  at $0.01 \mbox{ Hz}\leq f\leq 100$ Hz. Promising frequency peaks appearing in the power spectra 
  of both observations were cross-correlated and selected for subsequent searches using standard 
  epoch-folding analysis.

  PZT99 reported the detection of a periodic signal with $P=75.2797300$ ms and $\dot{P}=1.49
  \times10^{-13}\mbox{s s}^{-1}$ in their ROSAT analysis. We searched for coherent pulsations 
  in a period range extrapolated to these spin parameters in XMM-Newton data. The photon statistics 
  of this data, compared to the ROSAT data, is about a factor 25 higher. A similar analysis as 
  reported in the present work was already performed by Becker \& Aschenbach (2002) who could 
  not confirm the existence of a pulsed signal in an extrapolated period range, neither in the 
  ROSAT nor in the XMM-Newton data. We herewith fully confirm their result in our independent 
  analysis.

\begin{figure*}
\centerline{\psfig{figure=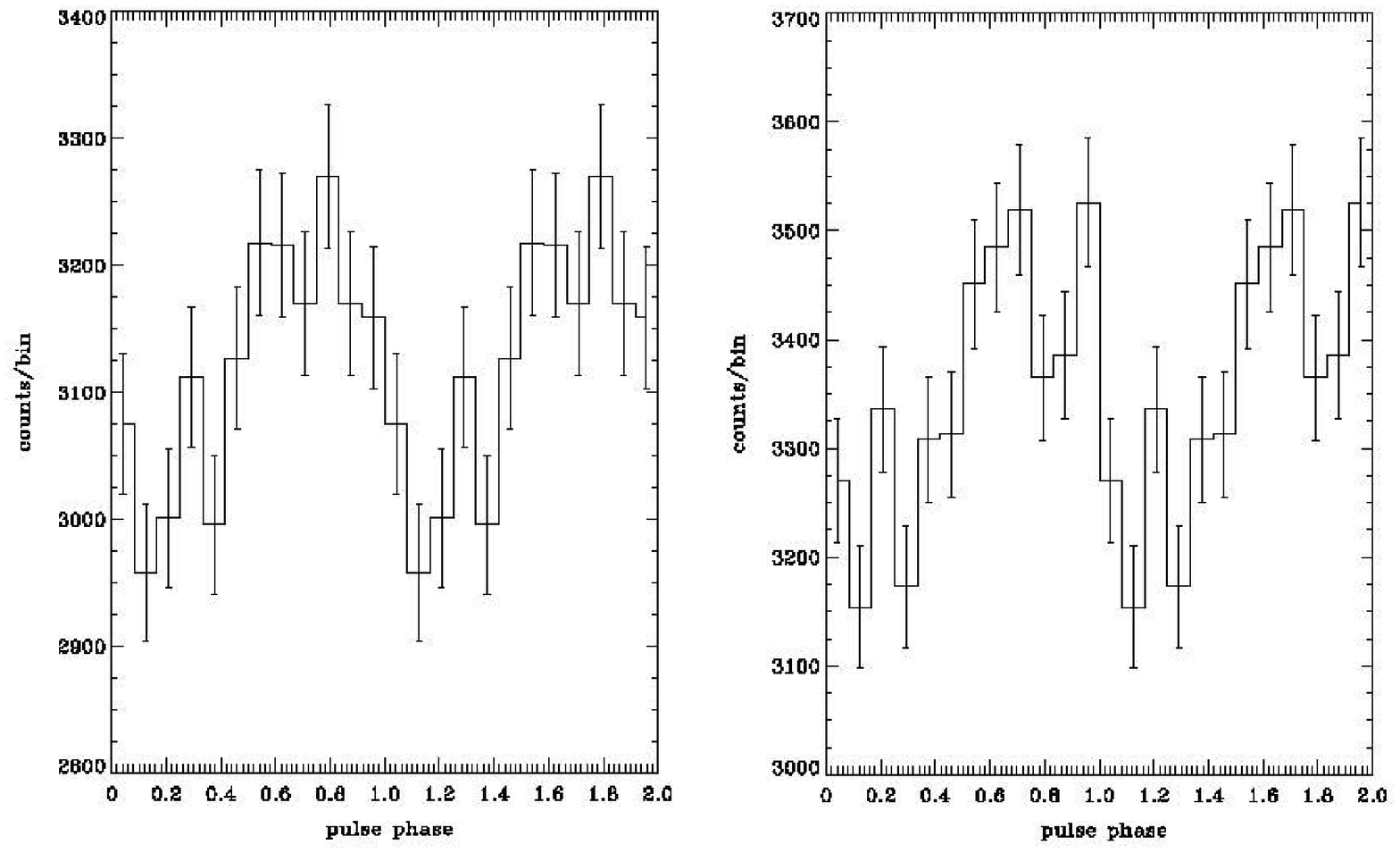,width=18cm,clip=}}
\centerline{\psfig{figure=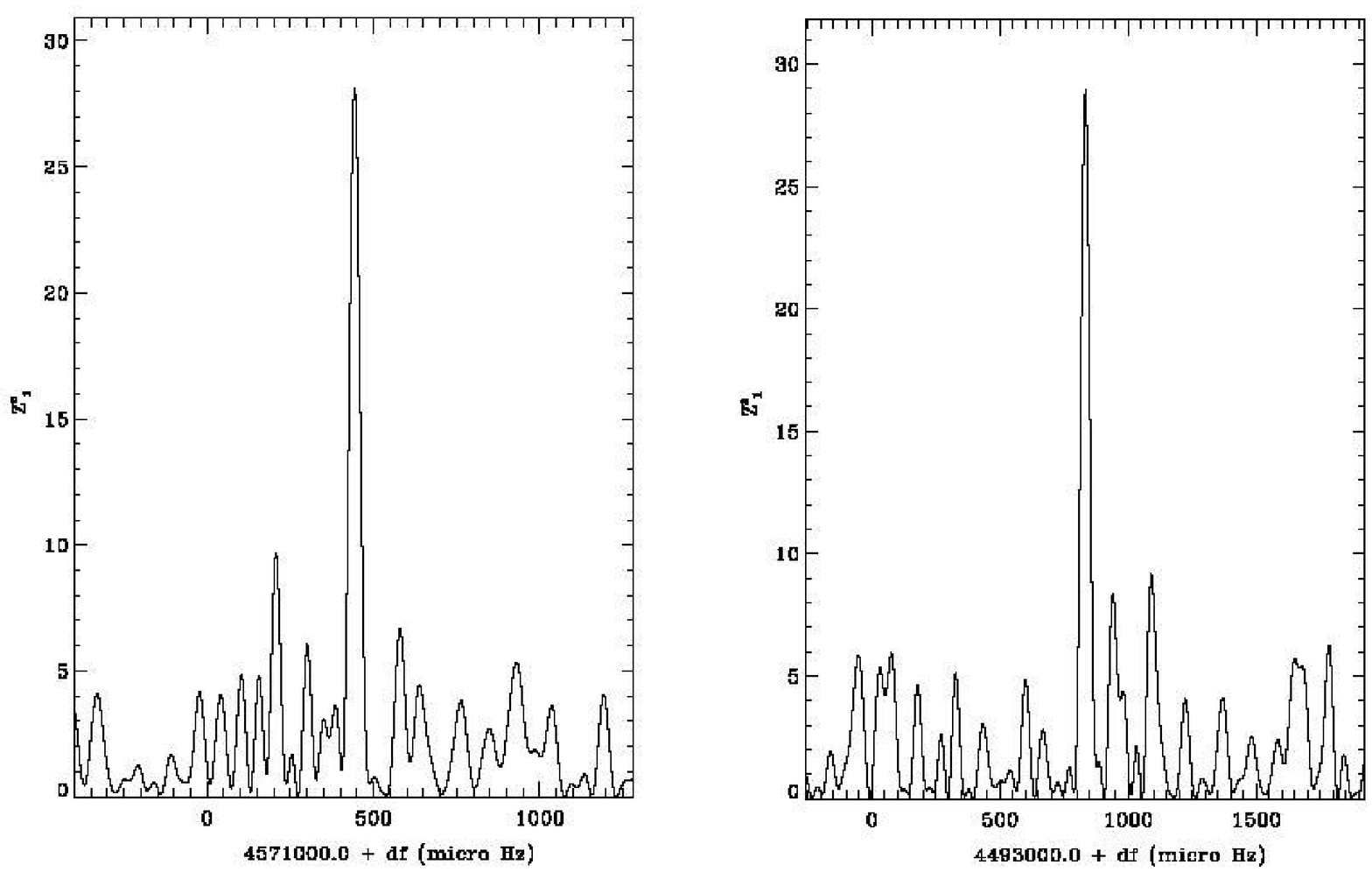,width=18cm,clip=}}
\caption{Pulse profiles and $Z^2_1$-distribution of the folded light curves as found in the XMM-Newton EPIC-PN data 
taken in April (left panel) and November 2001 (right panel).}
\label{pulses}
\end{figure*}

  Taking peaks in the power spectra as initial candidates, we made a more detailed search 
  using $Z^{2}_{m}$ test where $m$ is the numbers of harmonics (Buccheri et al.~1983). 
  We have detected periodicities of $P=0.218749\pm0.000001$ s in XMM1 (MJD 52014.4634390827268 days)
  \footnote{The mean epoch of the observation in TDB at the solar system barycenter (SSB)} and $P=0.222527\pm0.000002$ s 
  in XMM2 (MJD 52221.8938398198225 days) which both have very similar properties. The quoted uncertainties indicate the 
  Fourier resolution $P^{2}/T$ in the corresponding observation, where $T$ is the time 
  span in the data set. Using the $H$-test (De Jager, Swanepoel, \& Raubenheimer 1989), 
  we found that $H$ is maximized for the first harmonic. The calculated $Z^{2}_{1}$ for the 
  detected signals in XMM1 and XMM2 are 28.10 and 28.92, respectively.
  The nominal probabilities for the identification of these signals by 
  chance are $8\times10^{-7}$ and $5\times10^{-7}$, respectively. 
  The pulse profiles and $Z^{2}_{1}$ distribution is given in Figure \ref{pulses}.
  Both lightcurves are similar to each other and share the same sinusoidal
  characteristics. Following Becker \& Tr\"umper (1999), we calculated 
  the pulsed fraction of this signal by the bootstrap method 
  proposed by Swanepoel, de Beer, \& Loots (1996) and obtained 
  $P_f= 5\pm1\%$ in both, the XMM1 and XMM2 data sets. The period 
  time derivative calculated from the separation of the epochs of the 
  two data sets is $\dot{P}=(2.112\pm0.002)\times10^{-10}\mbox{s s}^{-1}$.

  Taking the number of $10^{5}$ trials into account, 
  the probabilities for finding these signals by chance is $(5-8)\times 10^{-2}$. 
  However, given the similarity in period, pulse shape, signal strength and pulse fraction  
  together with the detection of the signal in two independent XMM-Newton data 
  sets makes this signal a very promising candidate to test and search for in 
  future observations. 

  In order to minimize the probability of a false detection we have investigated 
  the possibility that the signals are induced from the readout processes in 
  the detector CCDs or other cyclic processes operating during data acquisition. 
  For this we have extracted events from
  \Pu\, in both, the XMM1 and XMM2 data sets from a location  near to \RX,
  i.e.~from CCD columns located at the same level in the readout direction 
  as \RX. The same procedure of timing analysis was applied on these events 
  as applied to the events from \RX. However, we did not detect any cyclic 
  signal at a period near to $\sim 0.22$ s. We therefore can rule out that 
  the detected pulsations are due to periodic systematics in the 
  on-board data processing during data acquisition.

  To further cross-check this periodicity detection, we utilized an independent 
  data set from Chandra HRC-S. Since the observation with HRC-S was performed 
  in ``imaging" mode where the outer segments of the micro-channel plate were 
  disabled, the total count rate is below the telemetry saturation limit, so 
  that all events can be assigned with accurate time and the HRC timing anomalies 
  are minimized. This enables us to perform an accurate timing analysis on this 
  data set, though the photon statistics is a factor of $\sim6$ lower than in
  the XMM-Newton data. The event file of HRC-S was firstly barycentric corrected 
  (with the position given in the HRC-I image fit)  by CIAO tool AXBARY. With the 
  $P$ and $\dot{P}$ estimated from XMM-Newton data, we extrapolated the period to 
  the epoch of HRC-S observation as an initial starting point. A detailed search 
  around this period gives a promising candidate at $P=0.217303\pm0.000002$ s in 
  the Chandra HRC-S data (MJD 51934.6266560833901 days). 
  The $H$-test indicates the highest probability for $Z^{2}_{7}=45.94$
  wich yields a nominal chance probability of $3 \times 10^{-5}$. This is not strong
  enough to conclude a significant signal in the HRC-S data though we point out that
  there are only $\sim6000$ counts available for this test. For a 5\% 
  pulsed fraction as indicated in the XMM-Newton data only $\sim300$ counts would 
  contribute to the pulsed component. The low significance of the signal found in 
  the Chandra data thus would be in line with a low significance of the periodic 
  signal. Archival ROSAT and ASCA data of \RX\, are of small photon statistics
  so that we did not include this data in order to search for a pulsed signal near
  to 0.22 s.

\section{Discussion \& Conclusion}

  The lack of any detectable long term variability together with the high X-ray/optical 
  flux ratio and the observed spectral characteristics makes it very unlikely that \RX\, 
  is something else but the compact stellar remnant formed in the core collapsed supernova 
  which was left behind  \Pu.  Our image analysis shows that the compact object is the 
  hardest X-ray source in the 30 arcmin central region of \Pu. Chandra and XMM-Newton 
  data do not show  any extended X-ray emission which could be plerionic emission powered 
  by the compact remnant, though this is difficult to quantify given the patchy environment 
  in which \RX\, is located.

  From the spectral fitting, we found that the point-source spectrum is compatible with 
  a two component blackbody model. The best-fit model yields $N_{H}=4.54^{+0.49}_{-0.43}
  \times10^{21}\mbox{cm}^{-2}$, temperatures of $T_{1}=2.61^{+0.30}_{-0.26}\times10^{6}$K 
  and $T_{2}=5.04^{+0.28}_{-0.20}\times10^{6}$K for the projected blackbody emitting 
  areas with radii $R_{1}=3.29^{+1.12}_{-0.74}$ km and $R_{2}=0.75^{+0.12}_{-0.15}$ 
  km, respectively. Compared to standard cooling curves (e.g.~Yakovlev et al.~2004) 
  $T_{1}$ is a little  higher than $\sim 1.8\times10^{6}$ K which would be expected 
  for a 1.35 $M_{\odot}$ neutron star with a stiff equation of state (Prakash et al.~1988). 
  $R_{1}$ is not quite consistent with the size of a typical neutron star. This was  
  already found by ZTP99 who attempted to obtain a set of reasonable neutron star parameters 
  by modeling the ROSAT data with spectral models which take the presence of a hydrogen 
  atmosphere into account, though the applicability of these models are restricted.

  In the present work we have shown that a parameter set which is consistent with a 
  standard cooling neutron star model can be obtained by fixing $R_{1}$ at 10 km. This
  yields a column density of $6.38^{+0.21}_{-0.13}\times10^{21}\mbox{cm}^{-2}$, 
  temperatures of $T_{1}=1.87^{+0.02}_{-0.02}\times10^{6}$K and 
  $T_{2}=4.58^{+0.03}_{-0.07}\times10^{6}$K for a projected blackbody emitting area with
  radius $R_{2}=1.09^{+0.04}_{-0.04}$ km.

  The results of our analysis suggest that the low temperature component is emitted from 
  a large fraction of the neutron star surface, while the high temperature component is 
  emitted from a much smaller and hotter region. This double blackbody model, though, could
  be a two-step adaption for a wider temperature distribution which centrally peaks. Such 
  a hot spot on the neutron 
  star surface can be produced by several mechanisms. One of them is the bombardment of 
  the polar cap regions by energetic particles accelerated in the magnetosphere 
  backwards to the neutron star surface (Cheng, Ho \& Ruderman 1986; Cheng \& Zhang 1999). 
  Another way to produce a hot spot on the stellar surface is by anisotropic heat transport 
  (Page 1995). Since the heat conduction inside a neutron star is much more efficient in 
  the direction along the magnetic field lines than that in the perpendicular direction, a 
  complete model of cooling magnetic neutron star should lead to an anisotropic heat flow 
  and hence produce hot spots on the stellar surface. In this scenario, one should expect 
  the emission to be pulsed at the rotation period of the star as the hot spot goes across 
  the line of sight. A pulsed X-ray flux as revealed by the putative periodic signals seen
  in XMM1 and XMM2 thus would support this scenario.

  The pulsed fraction of the putative periodic signal is $5\pm1\%$ in XMM1 and XMM2. The emission 
  from young pulsars like the Crab is compatible with being 100\% pulsed (Tennant et al.~2001), 
  whereas the fraction 
  of pulsed photons is $\sim 7\%$ for the Vela pulsar and $\sim 20-40\%$ in many of the other 
  X-ray detected pulsars (e.g.~Becker \& Pavlov 2001; Becker \& Aschenbach 2002; Kaspi et 
  al.~2004).  A low pulsed fraction, however, is not unexpected though. When the general 
  relativistic effect is taken into account (Page 1995; Hui \& Cheng 2004), the pulsations 
  are found to be strongly suppressed and the pulsed fraction is highly dependent on the 
  mass to radius ratio of the star, the orientation of the hot spot and the viewing angle 
  geometry. This is due to the fact that the gravitational bending of light will make more 
  than half of the stellar surface become visible at any instant and hence the contribution   
  of the hot spot will be hampered. If the orientation of the hot spot is deviated from
  that of an orthogonal rotator and/or the star has a high mass to radius ratio, then a 
  very low amplitude pulsations is expected, which makes the periodicity search difficult.

  The $\dot{P}$ deduced for the candidate periodicity would be among the largest 
  spin-down rates in the neutron star population. The largest known $\dot{P}$ was 
  inferred from SGR 1806-20, $\dot{P}=
  (8-47)\times10^{-11}$s s$^{-1}$, (Kouveliotou  et al.~1998; Woods et al.~2002). If 
  the identifications of $P$ and $\dot{P}$ are correct, it implies a non-steady spin-down 
  behavior of \RX. This phenomenon is not unobserved. There are two SGRs (SGR 1806-20 
  and SGR 1900+14) which show large changes in the spin-down torque up to a factor of 
  $\sim4$ (Woods et al.~2002). Moreover, deviations from a steady spin-down were also 
  observed in the other radio-quiet neutron stars such as in SNR PKS 1209-51/52 
  (Zavlin, Pavlov, \& Sanwal 2004). 
  However, the rotational dynamics cannot be determined without ambiguity here and 
  further observations are needed to confirm and/or refine this putative periodicity.

  An alternative proposal to explain the origin of X-rays from radio quiet compact objects
  in supernova remnants is accretion onto a neutron star (e.g.~Pavlov et al.~2000). In this 
  scenario, the observed luminosity $L$ is powered by an accretion rate of $\dot{M}=L/(\zeta c^{2})$. 
  $\zeta$ is the accretion efficiency which is expressed as $\zeta=0.2M_{1.4}R_{6}^{-1}$. 
  Equating the expression of $\dot{M}$ with Bondi formula (i.e.~$\dot{M}=4\pi G^{2}M^{2}\rho v^{-3}$), 
  we can express the relation of the circumstellar baryon density as $n=8\times10^{3}v_{100}^{3}(0.2/
  \zeta)M_{1.4}^{-2}L_{33}$ cm$^{-3}$, where $v_{100}$ is the velocity of the neutron star in 
  the unit of 100 km s$^{-1}$ and $L_{33}$ is the luminosity in the unit of $10^{33}$ erg s$^{-1}$. 
  The offset of $\sim6.1$ arcmin from the optical expansion center, estimated distance 
  ($\sim2.2$ kpc) and estimated age ($\sim3700$ years) suggest a transverse velocity of 
  $\sim985$ km s$^{-1}$. Even we take $L_{33}=1$, which is lower than the value inferred 
  from the acceptable spectral fit $L_{33}\sim5$, the expression of $n$ implies a density 
  of about 6 orders of magnitude higher than the expected value found by Petre et al.~(1982).

  It is therefore safe to reject the scenario that the observed X-rays are powered by 
  accretion from circumstellar matter. The stringent optical limit also rules out the 
  possibility that the accretion is from a massive companion. However, we cannot 
  completely exclude the possibilities that the central object is accreting from a 
  very close dwarf star or from a fossil disk (van Paradijs et al.~1995) which remained 
  after the supernova explosion. For the first possibility, even though such a compact 
  system is unlikely to remain bound in the disruption of the high mass progenitor 
  ($\gtrsim 25 M_{\odot}$ Canizares \& Winkler 1981), a deeper search for optical/IR 
  counterpart can still be beneficial. 

  As has been shown in this work, the properties of the compact object in \Pu\, are 
  similar to those of other radio quiet neutron stars in many aspects (e.g.~Cas A; 
  Pavlov et al.~2000; Chakrabarty et al.~2001; Murray et al.~2002). Even though the 
  nature of this class of object has not yet been completely resolved, their existence 
  has already revolutionized the conventional notion of neutron stars and their 
  environment. Since it is easier to detect and identify active radio/X-ray pulsars 
  than the radio-quiet neutron stars which are only observable in X-ray and located 
  in a patchy X-ray bright supernova remnant, it is plausible that they are more common than 
  canonical pulsars. It is not unlikely to assume that the relatively small number of 
  detected radio-quiet neutron stars is due to observational selection effects. In this 
  sense, identifying the nature of this class, including RX J0822-4300, is very important.

\begin{acknowledgements}
We acknowledge discussion with Bernd Aschenbach and thank J\"urgen Fath and our colleagues 
at MPE for their support. We also thank the referee for thoroughly reading the manuscript and 
provide us with many useful comments. 
\end{acknowledgements}

\clearpage
\begin{table}
\centering
\caption{Details of the XMM-Newton and Chandra observations of RXJ0822-4300 \label{obs_info}}
\begin{tabular}{llllclcl}
\\
\hline\hline
\\
Telescope & Instrument & Instrument Mode & Filter & Obs.ID    & Start Date  & Effective   & Net Rate\\
          &            &                 &        &           &             & Exposure   & (cts $\mbox{s}^{-1}$)\\
\\
\hline\\
Chandra    & HRC-I    & Default          & UVIS   & 749        & 1999-12-21 & 16 ks & 0.21$\pm$0.01\\ \\
Chandra    & ACIS-S 2 & Timed Exposure   & OBF    & 750        & 2000-01-01 & 11 ks & 0.71$\pm$0.01\\ \\
Chandra    & HRC-S    & Imaging          & UVIS   & 1851       & 2001-01-25 & 19 ks & 0.32$\pm$0.01\\ \\
XMM-Newton & MOS1     & PrimeFullWindow  & Medium & 0113020101 & 2001-04-15 & 7.3 ks & 0.50$\pm$0.01\\ \\
XMM-Newton & MOS2     & PrimeFullWindow  & Medium & 0113020101 & 2001-04-15 & 7.5 ks & 0.52$\pm$0.01\\ \\
XMM-Newton & PN       & PrimeSmallWindow & Thin   & 0113020101 & 2001-04-15 & 16  ks & 1.62$\pm$0.01\\ \\
XMM-Newton & MOS1     & PrimeFullWindow  & Thick  & 0113020301 & 2001-11-08 & 7.5 ks & 0.45$\pm$0.01\\ \\
XMM-Newton & MOS2     & PrimeFullWindow  & Thick  & 0113020301 & 2001-11-08 & 6.2 ks & 0.46$\pm$0.01\\ \\
XMM-Newton & PN       & PrimeSmallWindow & Thin   & 0113020301 & 2001-11-08 & 16  ks & 1.63$\pm$0.01\\
\\
\hline
\end{tabular}
\end{table}

\clearpage


\begin{landscape}
\begin{table}
\centering
\caption{Spectral Parameters inferred from  data from XMM1 and XMM2.\label{spec}}
\begin{tabular}{llllllllll}
\\
\hline\hline
\\
Parameter & BB  & BB+BB & BB+BB$^{a}$  & PL & BB+PL  & BB+PL$^{b}$  & BB+PL$^{c}$  & BKPL & BREMSS\\
\\
\hline
\\
	  $N_{H}$($10^{21}\mbox{cm}^{-2}$) & $2.67^{+0.09}_{-0.09}$ & $4.54^{+0.49}_{-0.43}$ & $6.38^{+0.21}_{-0.13}$ & $9.26^{+0.14}_{-0.17}$ & $8.57^{+0.88}_{-0.35}$ & $4.00$ & 8.02 & $7.62^{+0.17}_{-0.36}$ & $5.76^{+0.11}_{-0.11}$ \\
	  \\
	  $\Gamma_{1}$                     & - & - & - & $4.29^{+0.02}_{-0.02}$ & $4.67^{+0.14}_{-0.05}$ & $2.51^{+0.11}_{-0.13}$ & 3.50 & $3.61^{+0.05}_{-0.05}$ & -\\
	  \\
	  $\Gamma_{2}$                     & - & - & - & - & - & - & - & $5.29^{+0.20}_{-0.18}$ & -\\
	  \\
	  $T_{1}$ ($10^{6}$K)              & $4.32^{+0.04}_{-0.03}$ & $2.61^{+0.30}_{-0.26}$ & $1.87^{+0.02}_{-0.02}$ & - & $4.43^{+0.05}_{-0.07}$ & $3.80^{+0.06}_{-0.05}$ & 1.87 & - & $9.57^{+0.18}_{-0.06}$\\
	  \\
	  $T_{2}$ ($10^{6}$K)              & - & $5.04^{+0.28}_{-0.20}$ & $4.58^{+0.03}_{-0.07}$ & - & - & - & - & - & -\\
	  \\
	  $R_{1}$ (km)                     & $1.23^{+0.03}_{-0.02}$ & $3.29^{+1.12}_{-0.74}$ & 10 & - & $0.89^{+0.06}_{-0.08}$ & $1.59^{+0.06}_{-0.06}$ & 10 & - & -\\
	  \\
	  $R_{2}$ (km)                     & - & $0.75^{+0.12}_{-0.15}$ & $1.09^{+0.04}_{-0.04}$ & - & - & - & - & - & -\\
	  \\
	  $N_{\mbox{bremss}}$ ($10^{-2}$ cm$^{-5}$)  & - & - & - & - & - & - & - & - & $1.28^{+0.02}_{-0.02}$\\
	  \\
	  $F_{X_{0.1-2.4keV}}$ (ergs cm$^{-2}$ s$^{-1}$) & $5.82\times10^{-12}$ & $9.67\times10^{-12}$ & $2.04\times10^{-11}$ & $1.77\times10^{-9}$ & $2.60\times10^{-9}$ & $1.18\times10^{-11}$ & $2.18\times10^{-10}$ & $3.45\times10^{-10}$ & $2.40\times10^{-11}$\\
	  \\
	  $F_{X_{0.5-10keV}}$ (ergs cm$^{-2}$ s$^{-1}$) & $6.05\times10^{-12}$ & $8.94\times10^{-12}$ & $1.48\times10^{-11}$ & $4.46\times10^{-11}$ & $3.88\times10^{-11}$ & $8.28\times10^{-12}$ & $2.69\times10^{-11}$ & $2.48\times10^{-11}$ & $1.27\times10^{-11}$\\
	  \\
	  Reduced $\chi^{2}$              & 1.82 & 1.20 & 1.28 & 1.91  & 1.21 & 1.42 & 3.77 & 1.22 & 1.20\\
	  \\
	  D.O.F                           & 467  & 465  & 466 & 467 & 465 & 466 & 467 & 465 & 467\\
\\
\hline
\\
\\
\end{tabular}
\end{table}
The emitting areas inferred from the blackbody fits are calculated for the assumed distance of 2.2 kpc.\\ 
\indent $F_{X_{0.1-2.4keV}}$ and $F_{X_{0.5-10keV}}$ are unabsorbed fluxes in the energy range $0.1 - 2.4$ keV and $0.5 - 10$ keV, respectively.\\
\indent $^{\mathrm{a}}$ $R_{1}$ fixed at 10 km radius.\\
\indent $^{\mathrm{b}}$ $N_{H}$ fixed at $4\times10^{21}\mbox{cm}^{-2}$.\\
\indent $^{\mathrm{c}}$ $R_{1}$ and $T_{1}$ fixed at 10 km and $1.87\times10^{6}$ K respectively.\\
\end{landscape}



\begin{table*}
\centering
\caption{Observed fluxes inferred from observations at different epochs.\label{flux}}
\begin{tabular}{lccl}
\\
\\
\hline\hline
\\
Observations  & $F_{X}$ (0.1-2.4 keV) & $F_{X}$ (0.5-10 keV) & Start Date\\
              & ergs cm$^{-2}$ s$^{-1}$ & ergs cm$^{-2}$ s$^{-1}$ & \\
\\
\hline\\

XMM1         & $3.37^{+0.28}_{-0.26}\times10^{-12}$ & $4.19^{+0.39}_{-0.36}\times10^{-12}$   & 2001-04-15\\[2ex] 
XMM2         & $3.38^{+0.30}_{-0.28}\times10^{-12}$ & $4.17^{+0.42}_{-0.37}\times10^{-12}$   & 2001-11-08\\[2ex] 
Chandra ACIS & $3.34^{+0.40}_{-0.36}\times10^{-12}$ & $4.18^{+0.54}_{-0.50}\times10^{-12}$   & 2000-01-01\\[2ex]
ROSAT PSPC $^{\mathrm{a}}$   & $\sim3\times10^{-12}$  &  -  & 1991-04-16\\
	    \\
\hline
\\
\end{tabular}
\\
$^{\mathrm{a}}$ Flux observed by ROSAT according to PBW96
\end{table*}


\clearpage

\clearpage

\clearpage

\clearpage

\clearpage

\clearpage

\clearpage

\clearpage

\clearpage


\clearpage

\end{document}